# D$^2$RLIR : an improved and diversified ranking function in interactive recommendation systems based on deep reinforcement learning

Vahid Baghi[1]     Seyed Mohammad Seyed Motehayeri[1]     Ali Moeini[1]     Rooholah Abedian[1]

[1] Dept. of algorithms and computation, School of Engineering Science, University of Tehran
{vahid.baghi, motehayeri, moeini, rabedian}@ut.ac.ir

**Abstract**

Recently, interactive recommendation systems based on reinforcement learning have been attended by researchers due to the consider recommendation procedure as a dynamic process and update the recommendation model based on immediate user feedback, which is neglected in traditional methods. The existing works have two significant drawbacks. Firstly, inefficient ranking function to produce the Top-N recommendation list. Secondly, focusing on recommendation accuracy and inattention to other evaluation metrics such as diversity. This paper proposes a deep reinforcement learning based recommendation system by utilizing Actor-Critic architecture to model dynamic users' interaction with the recommender agent and maximize the expected long-term reward. Furthermore, we propose utilizing Spotify's ANNoy algorithm to find the most similar items to generated action by actor-network. After that, the Total Diversity Effect Ranking algorithm is used to generate the recommendations concerning relevancy and diversity. Moreover, we apply positional encoding to compute representations of the user's interaction sequence without using sequence-aligned recurrent neural networks. Extensive experiments on the MovieLens dataset demonstrate that our proposed model is able to generate a diverse while relevance recommendation list based on the user's preferences.

## 1. Introduction

The majority of conventional recommender systems consider the recommendation procedure as a static process. In other words, the recommender system has no immediate interaction with the users and cannot improve the recommendation model based on the users' immediate and long-term feedback. On the other hand, an interactive recommender system can modify its model to the users' preferences. Some works [1]–[5] formulated the interactive recommendation procedure as a multi-armed bandit problem. These works suppose that users' preferences would be left unchanged, but it's evident that the users' preferences may change over the time. For example, the users' interest may vary depending on their mood or temper. Due to the recent advancement in deep reinforcement learning and outstanding success in various tasks such as playing games [6]–[8], robotics [9], [10], and other sequential decision-making problems [11], many research studies target reinforcement learning based recommender systems. Most existing reinforcement learning approaches are not reasonable in large discrete items space, which is the nature of recommender systems. In particular, traditional reinforcement learning techniques such as Q-learning [12] and Partially Observable Markov Decision Processes (POMDP) [13] are early efforts to model recommendation procedure as a reinforcement learning problem. However, these methods become impracticable with the growing number of items in the recommendation process. The conventional Deep Q-Network (DQN) based techniques [14] utilize a deep neural network as a function approximator to estimate Q-values for all available items in action space to take the best action with maximum Q-value in the current state. Although these techniques do not require to store Q-values in memory, the time complexity of Q-value evaluation to determine the best action is linear to the size of the action space and therefore becomes inefficient where the number of items is huge. In recent years, some works [15]–[17] proposed to leverage Actor-Critic architecture, which utilizes deep neural networks as separate policy and value function estimators. The policy network is known as the Actor, and the value network (i.e., Q-network) is known as Critic. In this architecture, actor-network generates a continuous parameter vector as action used to rank candidate items. Then, the critic evaluates the action generated by the actor using the Temporal Difference (TD) error. Despite solving temporal complexity in these studies [15]–[17], there are still two limitations: Firstly, a ranking function is applied to all items in action space used to produce recommendations by selecting items with the highest score. Therefore, the temporal complexity problem remains. Secondly, these methods [15]–[17] focus on recommendation accuracy. Hence, the system may offer only a narrow range of user's interests, such as only offering sci-fi movies, which is known as overfitting problem. In this paper, to tackle these problems, we propose a model to <u>d</u>iversifying in <u>d</u>eep <u>r</u>einforcement <u>l</u>earning-based <u>i</u>nteractive <u>r</u>ecommendation systems, namely D$^2$RLIR, by utilizing Actor-Critic architecture to model an interactive recommendation system. More specifically, to capture user's sequential behaviors, we add positional encodings [18] to the user's state embeddings (i.e., the latest n items of user's interaction history) and then, we fed user's state to the actor-network, which generates a continuous parameter vector as a proto-action. This proto-action is used to find similar items using the ANNoy algorithm [19]. Finaly, the TDE algorithm [20] is used to select Top-N items with the highest diversity. The critic-network estimates

Q-value (i.e., the maximum future reward of the proto-action in the current state) which is used to update actor and critic network weights. Extensive experiments on the MovieLens dataset demonstrate that the diversity of recommendation is increasing. The major contributions of this paper are in three folds :

- We propose an interactive recommendation framework based on deep reinforcement learning, which generates the recommendation list concerning both relevancy and diversity.
- In order to reduce redundant computation in generating the recommendation list, we propose to find similar items to the generated proto-action by actor-network using the ANNoy algorithm and then select Top-N items with the highest diversity.
- Instead of using RNN to capture user's sequential behaviors, positional encoding to compute representations of the user's interaction sequence proposed.

## 2. Related Works

### 2.1. Reinforcement Learning

Shani et al. [13] utilized a Markov decision process (MDP) formulation to model sequential decision problems in recommender systems using a predictive n-gram model for the initial MDP. Zhao et al. [16] proposed a novel list-wise recommendation model based on deep reinforcement learning (i.e., LIRD), which models the interactive recommendation system as a Markov Decision Process. Hu et al. [21] proposed to formulate a ranking control problem in E-commerce searching using the search session markov decision process (SSMDP) and leveraged the deterministic policy gradient (DPG) method to learn an optimal ranking policy. Zheng et al. [22] proposed a DQN-based reinforcement learning model for personalized news recommendation, which explicitly model future rewards and utilizes an effective exploration strategy to improve the diversity of the recommendation. Liu et al. [15] proposed a novel deep reinforcement learning based framework, named DRR, which explicitly model user-item interactions.

### 2.2. Diversity

The maximal marginal relevance (MRR) [23] was one of the earliest works to maximize relevance and diversity in re-ranking retrieved items used in text summarization. Ziegler et al. [24] proposed a topic diversification method to increase the diversity of recommendation list by decreasing the intra-list similarity (ILS). Bridge et al. [25] measured the distance between items using Hamming Distance, which enhanced the diversity of the recommendations by considering collaborative data only. To maximizing the diversity of the recommendation list while maintaining proportional similarity to the user request, Zhang et al. [26] modeled these goals as a bi-criterion optimization problem. Lathia et al. [27] showed the importance of temporal diversity in recommender systems by proposing an approach to measuring how the diversity of collaborative filtering data changes over the time. Abbassi et al. [28] proposed a (partition) matroid constraint algorithm to increase the diversity of representative documents set by avoiding selecting duplicate categories. Premchaiswadi et al. [20] proposed a ranking method named Total Diversity Effect Ranking (TDE) that increases the diversity of recommendation list by considering the overall diversity effect of each item. Lee et al. [29] constructed a strongly-connected and undirected graph that each node is an item, and each edge represents positive interactions between items. Then utilizing the concept of entropy, find recommendations by considering novelty and relevancy. Cheng et al. [30] modeled the diversification problem in the recommendation system as a supervised learning task and formulated this diversified recommendation task as two optimization problems. Then proposed a diversified collaborative filtering method to solve the optimization problems. Wang et al. [31] proposed a novel method named DivRec LSH to produce diversified recommendations, which was achieved by using Locality-Sensitive Hashing (LSH) technique. Liu et al. [32] proposed the determinantal point process (DPP) method to promoting diversity in recommendations.

## 3. Proposed Framework

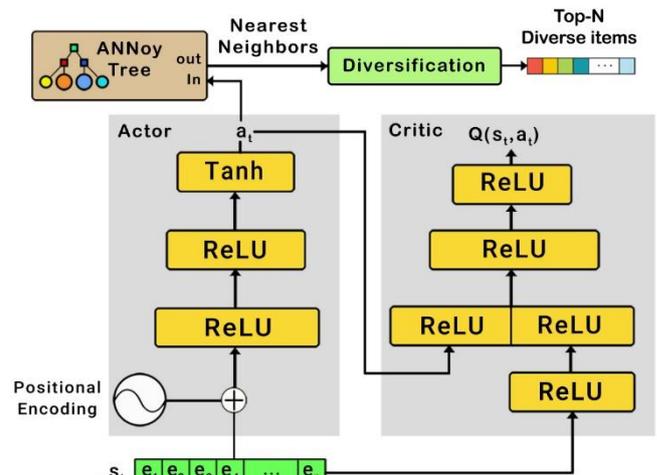

Fig. 1. The illustration of the proposed framework

### 3.1. Problem Statement

The recommender (i.e., agent) suggests a list of items with regard to both relevancy and diversity to the user (i.e., environment). Then, the user chooses one of them. The recommender agent performs this process sequentially over the time steps to maximize its cumulative reward. We model the recommendation process as a Markov Decision Process (MDP), which widely adopted in reinforcement learning. An MDP is a Markovian process described as follows :

- **State Space:** a state $s_t$ is defined as embeddings of latest user's positive interaction before time step $t$.
- **Action Space:** action $a_t \in \mathbb{R}$ is a continuous parameter vector that used to construct a diverse recommendation list.
- **Reward:** based on the chosen action at in state $s_t$

by the recommender agent, a list of diverse while relevant items is recommend to the user. Then, the user interacts with these items and provides him feedback, e.g., rating, click or purchase, etc. and the recommender receives immediate reward $R(s_t, a_t)$ according to diversity of recommendation list and the user's feedback.

- **Transition:** The user's state changes from $s_t$ to $s_{t+1}$, when the recommender agent takes action at at time step $t$. If the user ignores the whole recommended list, then the next state remains unchanged (i.e., $s_t = s_{t+1}$). Otherwise, the next state $s_{t+1}$ updates based on positive user's feedback.
- **Discount rate:** $\gamma \in [0,1]$ is a factor that determines the importance of future rewards relative to immediate reward. If $= 0$, the agent will be short-sighted by only considering immediate rewards. On the contrary, if $\gamma = 1$, the agent prefers long-term rewards. Hence, the discounted expected cumulative reward at time step $t$ denoted as $G_t = \sum_{k=0}^{T} \gamma^k R(s_t, a_t)$.

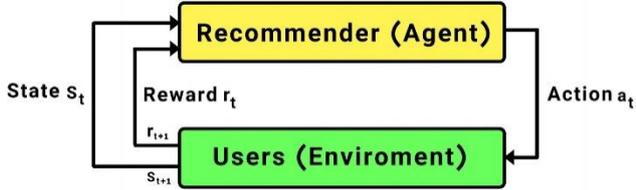

Fig. 2. the recommender-users interactions in MDP

The general scenario of the recommender-users interactions in MDP is illustrated in Fig. 2. In each state $s_t$, the recommender (i.e., agent) chooses a proto-action $a_t$ based on current policy $\pi$. This proto-action is a continuous parameter vector which is used to generate a list of items. Then, the recommender updates the policy $\pi$ based on user's feedback $r_{t+1}$ and user's state changes from $s_t$ to $s_{t+1}$. The objective of the reinforcement learning is to find a recommendation policy $\pi : S \to A$ to maximize the expected reward.

### 3.2. The Actor Network

The actor-network utilizes a deep neural network to learn the optimal policy, i.e., generating the best action $a_t$ based on the current user's state $s_t$. The user's state can be defined as $s_t = \{e_i | e_i \in \mathbb{R}^m, 1 \leq i \leq n\}$ such that $e_i$ is embedding[1] of $i^{th}$ item in the latest $n$ user's positive interaction history. When the recommender agent generates a list of items and then recommends them to the user, if the user skips whole items in the list, then the next state remains unchanged, i.e., $s_t = s_{t+1}$. If $p = \{p_i | p_i \in \mathbb{R}^m, 1 \leq i \leq r \leq n\}$ are recommended items that the user has given positive feedback to them, then the next state $s_{t+1}$ updates as follows:

$$s_{t+1} = \{e_i | e_i \in s_t, r + 1 \leq i \leq n\} \cup \{p_i | p_i \in p, 1 \leq i \leq r\} \quad (1)$$

such that chronological order of items in $s_{t+1}$ is preserved. Since the current representation of items in the user's state has no sense of how the items are in chronological order, we add Positional Encoding to each item's embedding to determine the position of each item in the sequence. The positional encoding formula is as follows:

$$PE(pos, i) = \sin\left(\frac{pos}{10000^{\frac{2i}{d_{model}}}}\right) \quad (2)$$

$$PE(pos, 2i + 1) = \cos\left(\frac{pos}{10000^{\frac{2i}{d_{model}}}}\right) \quad (3)$$

For example, if $s_t = \{e_1, \ldots, e_{pos}, \ldots, e_n\}$ is the user's state, then positional encoding operation of $e_{pos} \in \mathbb{R}^m$ with $d_{model} = m$, would be as below:

$$PE(e_{pos}) = e_{pos} + \left[\sin\left(\frac{pos}{10000^{\frac{(2\times 0)}{m}}}\right), \cos\left(\frac{pos}{10000^{\frac{(2\times 0)}{m}}}\right),\right.$$
$$\left.\ldots, \sin\left(\frac{pos}{10000^{\frac{\left(2\times\frac{m-2}{2}\right)}{m}}}\right), \cos\left(\frac{pos}{10000^{\frac{\left(2\times\frac{m-2}{2}\right)}{m}}}\right)\right] \quad (4)$$

the output of the positional encoding step fed into two fully-connected ReLU layers and one Tanh layer that maps the current state $s_t$ to a proto-action at in this way:

$$f_{\theta^\pi}(s_t) = a_t \quad (5)$$

where the function $f_{\theta^\pi}$ parameterized with $\theta^\pi$, mapping from the state space $\mathbb{R}^{1 \times mn}$ to the action space $\mathbb{R}^m$. The generated proto-action by the actor-network is not a valid action; i.e., it may not exist in the item space. Hence, we use the approximate nearest neighbor search (ANNS) to find the closest items to proto-action $a_t$, and then generate a diverse list using TDE algorithm.

### 3.3. Diversification

Given generated proto-action by the actor-network, a list of items is recommended to the user. Unlike the existing ranking functions [15]–[17], [33], which compute the distance from the proto-action to every other item in the action space to find similar items, we find κ-similar items to the proto-action using Approximate Nearest Neighbors Oh Yeah (ANNoy) [19], which achieved high performance in existing benchmarks [34], [35]. The ANNoy algorithm builds a binary tree where each intermediate node is a random hyperplane chosen by picking two random points in space and then splitting space by the hyperplane equidistant between them. This process is done recursively in each sub-space until each node contains at most $\nu$ items. The hyperplanes of first 400 items in the Movielens-100K dataset shown in Fig. 3a and corresponding binary tree shown in Fig. 3b.

---

[1] In this paper, we use a probabilistic matrix factorization (PMF) technique as a embedding model

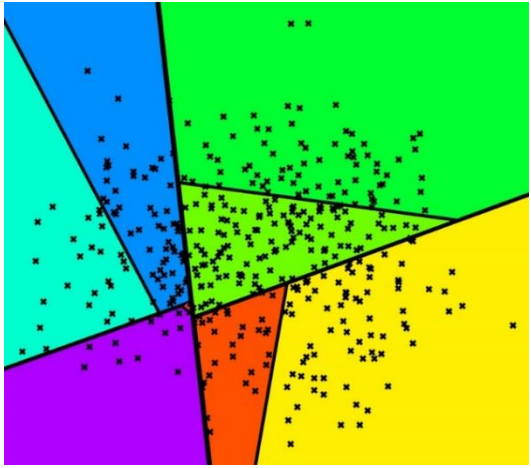

(a) Hyperplanes

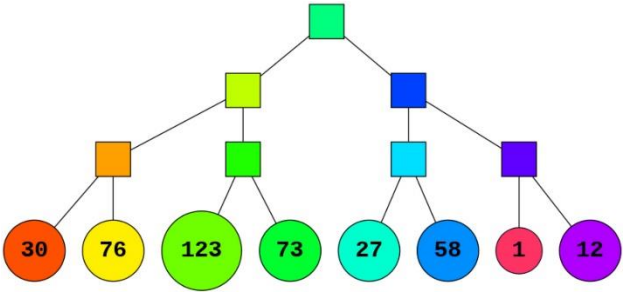

(b) Binary Tree

Fig. 3. An example of ANNoy algorithm

Once the binary tree has been built, we find $k$-nearest points for the generated proto-action $a_t$ by traversing the binary tree from root to leaf. Since each leaf (i.e., sub-space) has at most $v$ points, it is may be $v \leq k$. To overcome this problem, [19] proposed to construct multiple trees (i.e., a forest) and combine search results from all trees. The roots of these trees are stored in a priority queue, and the search continues until $k$ candidates are collected. In order to generate a diverse recommendation list, after removing duplicate candidates and sorting remaining items in ascending order of angular distance, we store candidate items in $C_{s_t}$ and then compute the Total Diversity Effect (TDE) of each item $c_i \in C_{s_t}$ as follows :

$$TDE(c_i) = \sum_{j=1}^{|C_{s_t}|}(1 - \frac{c_i \cdot c_j}{||c_i|| \cdot ||c_j||}) ; i \neq j \ \& \ c_i, c_j \in C_{s_t} \quad (6)$$

Finaly, we sort candidate items $C_{s_t}$ in descending order of TDE and select Top-N items as recommendation list. The pseudo code of this process is shown in Algorithm 1.

---

**Algorithm 1 :** Generate diverse recommendation list

**Input:** $a_t$ ; generated proto-action by actor-network
**Output:** $d_t$ diverse recommendation list
1: find $k$-similar items to $a_t$ and store in $C_{s_t}$
2: **for** $i = 1$ to $|C_{s_t}|$ **do**
3:    Compute $TDE(c_i)$ usnig Eq. 6
4: **end for**
5: $candidates$ = sort $C_{s_t}$ in descending order of TDE
6: $d_t$ = select Top-N items from $candidates$
7: **return** $d_t$

---

In Zhao et al. [16], in order to generate a list of recommendations of size $k$, the actor-network generates $k$ weight vectors of length $N$, which $N$ is the size of the item space; hence the computational complexity is $O(kN)$. In [15], the generated proto-action by actor-network is used to rank all items in item space; in consequence, the computational complexity is $O(N \log N)$. In [32], in order to generate recommendations, a fast greedy MAP inference algorithm is proposed, whose computational complexity is $O(k^2 N)$. The computational complexity of our proposed method for generating a recommendation list is $O(k \log N)$. The computational complexity of each query in the ANNoy algorithm is $O(\log N)$. In the worst case, if only one item is selected in each query, the required number of queries is $k$, so the computational complexity of our proposed method is $O(k \log N)$.

### 3.4. The Critic Network

As shown in Fig.1, the critic network inputs (state $s_t$, proto-action $a_t$) pair and outputs the estimation of the state-action value function $Q(s,t)$, namely Q-function. The Q-function denotes the goodness of generated action $a_t$ in state $s_t$ by following a policy $\pi$. The critic network utilizes a deep Q-network (DQN) with parameter $\theta^Q$, which uses a deep neural network as a function approximator. Therefore, The Q-function that specifies the expected return from state $s_t$ with taking an action $a_t$ under policy $\pi$, can be defined as follows:

$$Q(s_t, a_t \mid \theta^Q) = E_{s_{t+1}}[r_t + \gamma Q(s_{t+1}, a_{t+1} \mid \theta^Q)] \quad (7)$$

### 3.5. Reward Function

The recommender receives immediate reward $R(s_t, a_t)$ according to the diversity of the generated recommendation list and the user's feedback. The reward function directly affects the result of Eq.7, which estimates the state-action value function $Q(s_t, a_t)$. Therefore, we need to ensure the performance of the reward function before training the algorithm. In this paper, we define the reward function as follows:

$$R(s_t, a_t) = \text{Diversity}(t) \cdot \text{Rating}(t) - \lambda \quad (8)$$

The function $Diversity(t)$ is the intra-list distance (ILD) [26] of recommended items $d_t$, which defined as below :

$$\text{Diversity}(t) = \frac{2}{|d_t|(|d_t|-1)} \sum_{i_l, i_k \in d_t} distance(i_l, i_k) \quad (9)$$

and the function $Rating(t)$ is the average user rating to recommended items $d_t$, which defined as below :

$$\text{Rating}(t) = \frac{\sum_{i \in d_t} r(i)}{|d_t|} \quad (10)$$

where $r(i)$ is current user's rating to item $i$, which simulated by pre-trained PMF model [36]. The parameter λ determines when the agent receives a negative reward. The agent will always receive positive rewards and not be penalized if this parameter does not exist. As shown in Fig. 4, we can plot

$Rating(t)$ in terms of $Diversity(t)$ to show how the reward function changes to improve $Diversity(t)$ and $Rating(t)$. The ideal goal in this plot is to get to the top right corner, i.e. $Diversity(t) = 1$ and $Rating(t) = 5$, which our reward function changes smoothly to achieve this goal.

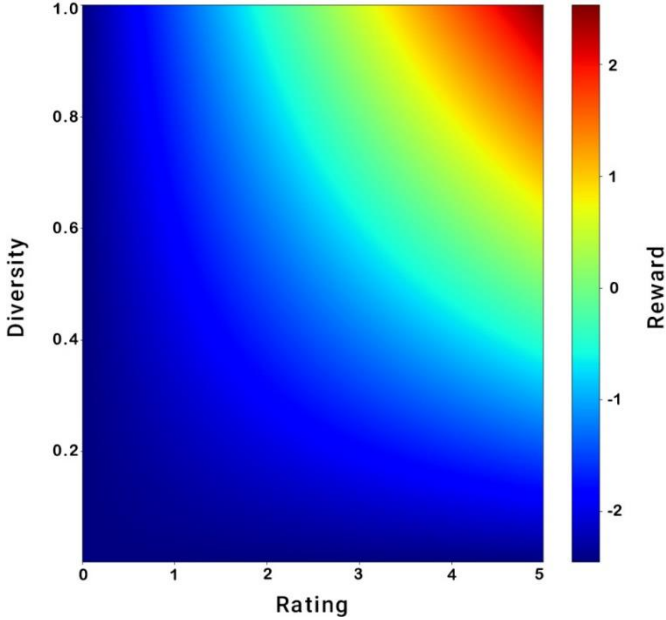

Fig. 4. The illustration of the reward function

## 3.5. Training

In this paper, we use the Deep Deterministic Policy Gradient (DDPG) algorithm [37] to train our proposed model. In DDPG, there are four neural networks which are as follows: 1) a critic network with parameters $\theta^Q$ 2) an actor-network with parameters $\theta^\mu$ 3) a target critic network with parameters $\theta^{Q'}$ 4) a target actor-network with parameters $\theta^{\mu'}$. In each episode, a user's state is chosen, and the recommender interacts with the user in $T$ step. In each step, the actor-network generates a proto-action at to construct a diverse list $d_t$ using Algorithm 1. Then, the user provides him feedback to recommendation list $d_t$, and the recommender receives immediate reward $r_t$. The user's state is updated according to Eq. 1. After that, a tuple $(s_t, a_t, r_t, s_{t+1})$ stores in the replay buffer. In order to update the critic network $Q(s_t, a_t|\theta^Q)$, a mini-batch of size $N$ is sampled from experience replay $R$ to calculate the temporal Temporal difference (TD) target value :

$$y_i = r_i + \gamma Q'\big(s_{i+1}, \mu'(s_{i+1} \mid \theta^{\mu'}) \mid \theta^{Q'}\big) \qquad (11)$$

Then, the network parameter $\theta^Q$ is updated by minimizing following the loss function :

$$\text{loss} = \frac{1}{N}\sum_i \big(y_i - Q(s_i, a_i \mid \theta^Q)\big)^2 \qquad (12)$$

Also, the objective of the actor-network is to maximize the expected return, which is defined as follows :

$$J(\theta) = E[Q(s_t, \mu(s_t, \theta^\mu) \mid \mu)] \qquad (13)$$

In order to optimize this objective function, the gradient of $J(\theta)$ respect to policy parameters $\theta^\mu$ can be taken using chain rule as follows:

$$\nabla_{\theta^\mu} J(\theta) = E_\mu[\nabla_a Q(s_t, a_t \mid \theta^\mu) \cdot \nabla_{\theta^\mu}(s_t \mid \theta^\mu)] \qquad (14)$$

Since the actor-network updates by sampled mini-batch of size $N$ from experience replay, the mean of $\nabla_{\theta^\mu} J(\theta)$ over $N$ samples is considered :

$$\nabla_{\theta^\mu} J(\theta) \approx \frac{1}{N}\sum_i [\nabla_a Q(s_i, \mu(s_i) \mid \theta^Q)\nabla_{\theta^\mu}\mu(s_i \mid \theta^\mu)] \qquad (15)$$

The target target networks are updated by a soft update technique as follows :

$$\begin{aligned}\theta^{Q'} &\leftarrow \tau\theta^Q + (1-\tau)\theta^{Q'}\\ \theta^{\mu'} &\leftarrow \tau\theta^\mu + (1-\tau)\theta^{\mu'}\end{aligned} \qquad (16)$$

where $\tau \ll 1$

the training algorithm based on DDPG is described in Algorithm 2.

---
**Algorithm 2 :** Training Algorithm of proposed model
---
1: Randomly initialize the network $f_{\theta^\pi}$ and critic network $Q(s, a \mid \theta^\mu)$ with the parameters $\theta^\pi$ an d $\theta^\mu$
2: Initialize target network $f'_{\theta^\pi}$ and $Q'(s, a \mid \theta^\mu)$ with weights $\theta^{\pi'} \leftarrow \theta^\pi$ and $\theta^{\mu'} \leftarrow \theta^\mu$
3: Initialize replay buffer $R$
4: **for** episode $= 1$ to $M$ **do**
5:    Initialize state $s_0 = [e_i, ..., e_n]$
6:    Add positional encoding to each $e_i \in s_0$ according to Eq. 4
7:    **for** $t = 1$ to $T$ **do**
8:      Generate action $f_{\theta^\pi}(s_t) = a_t$ according to current policy $\pi$
9:      Generate diverse list $d_t$ according to Algorithm 1
10:     Recommend list $d_t$ to the user
11:     Calculate reward $r_t$ according to Eq. 8
12:     Update user's state from $s_t$ to $s_{t+1}$ according to Eq. 1
13:     Store transition $(s_t, a_t, r_t, s_{t+1})$ in $R$
14:     Set $s_t \leftarrow s_{t+1}$
15:     Sample a mini-batch of N transitions $(s_i, a_i, r_i, s_{i+1})$ from $R$
16:     Set $y_i = r_i + \gamma Q'\big(s_{i+1}, \mu'(s_{i+1} \mid \theta^{\mu'}) \mid \theta^{Q'}\big)$
17:     Update the Critic network by minimizing the loss:
18:     $\text{loss} = \frac{1}{N}\sum_i \big(y_i - Q(s_i, a_i \mid \theta^Q)\big)^2$
19:     Update the Actor network using the sampled policy gradient :
20:     $\nabla_{\theta^\mu} J(\theta) \approx \frac{1}{N}\sum_i [\nabla_a Q(s_i, \mu(s_i) \mid \theta^Q)\nabla_{\theta^\mu}\mu(s_i \mid \theta^\mu)]$
21:     Update the target networks:
22:     $\theta^{Q'} \leftarrow \tau\theta^Q + (1-\tau)\theta^{Q'}$
23:     $\theta^{\mu'} \leftarrow \tau\theta^\mu + (1-\tau)\theta^{\mu'}$
24:    **end for**
25: **end for**

# 4. Experiments

## 4.1. Dataset

We use MovieLens datasets to evaluate our proposed framework. The MovieLens dataset contains a set of users' expressed rating for movies which collected by the MovieLens website[2]. In each row $i$, there is a tuple $(u_i, m_i, r_i, t_i)$ where $r_i$ is rating (in range [1-5]) of user $u_i$ to movie $m_i$ at a particular time $t_i$. **MovieLens-1M**[3] includes 1 million rating from 6040 users to 3706 movies. Since each user's state $s_t$ contains only positive interactions, we remove ratings less than 3 from both datasets.

## 4.2. Methods Comparison

We utilize following methods to compare performance of our proposed model:
- **LIRD [16]:** Proposed a list-wise recommendation model based on deep reinforcement learning that the actor-network generates a list of weight vectors and then scores all items in item space using these weights.
- **DRR-ave [15]:** Proposed a state representation module to model user-item interactions explicitly in deep reinforcement learning-based recommendation system. The generated proto-action by actor-network is used to rank all items in item space. Then, the recommender agent recommends an item with the highest score to the user.
- **D²RL [32]:** Proposed a fast greedy MAP inference algorithm to promoting diversity in generated recommendations.

In each dataset, we consider 80% of users' positive interactions for training and 20% of remaining for testing. Since the users' ratings to all movies are not available in datasets, we use pretrained PMF, in which each user and item is represented as 100-dimensional embeddings, to simulate unknown ratings. In DRR-ave method, the scoring function recommends one item with the highest score to the user. Hence, we select N items with the highest score to generate a list of items. The major hyperparameters are listed in Table 1. We employ **NDCG@k**, **Diversity@k** and **Precision@k** metrics to evaluate the performance of the recommendation models. In this paper, we define **Precision@k** as follows:

$$\text{Precision@N} = \frac{\sum_{i=1}^{N} r_{ui}}{N} \quad (17)$$

such that $r_{ui}$ is simulated rating of the user $u$ to the $i^{th}$ item in recommended list $d_t$ and $r_{ui} \geq 3$.

Table 1. The Major Hyperparameters

| Method | Major Hyperparameters |
|---|---|
| $D^2RLIR$ | $N = |d_t| = 10$ <br> $n = |s_t| = 10$ <br> $|C_{s_t}| = 30$ <br> $T = [1,10]$ <br> $\# tree = 5$ <br> $\lambda = 1.8$ <br> $d_{model} = 100$ |
| $DRR - ave$ | $N = |d_t| = 10$ <br> $n = |s_t| = 10$ <br> $T = [1,10]$ |
| $LIRD$ | $N = |d_t| = 10$ <br> $n = |s_t| = 10$ <br> $T = [1,10]$ |
| $D^2RL$ | $N = |d_t| = 10$ <br> $n = |s_t| = 10$ <br> $T = [1,10]$ |

## 4.3. Results

The evaluation results are summarized in Fig.5 to 8. The results show that the proposed model is better than other methods in 90% of cases in terms of diversity, 40% in terms of precision and 30% in terms of nDCG. In other cases, however, it is slightly different from other methods. Fig.8 indicates that the use of positional encoding improved diversity by 9%, precision by 8%, and nDCG by 0.5%. Therefore, complex structures such as CNN and RNN can be replaced by positional encoding.

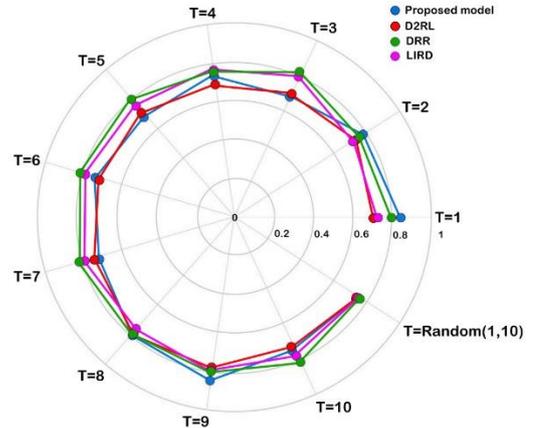

Fig. 5. Comparison of algorithms in terms of nDCG@10

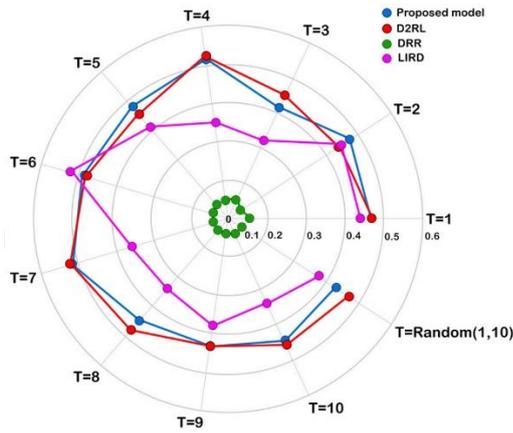

Fig. 6. Comparison of algorithms in terms of Precision@10

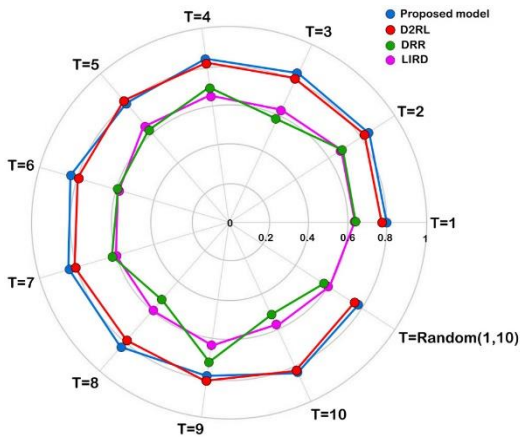

Fig. 7. Comparison of algorithms in terms of Diversity@10

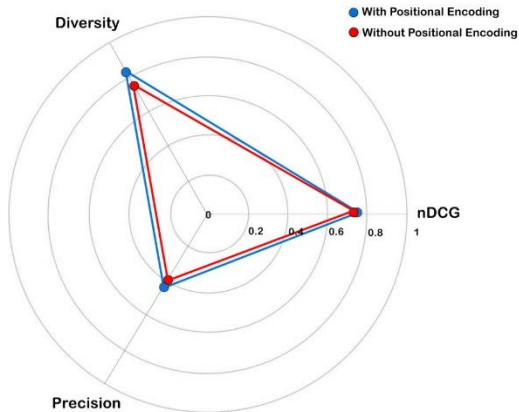

Fig. 8. Effect of positional encoding on results

## 5. Conclusion and Future Work

In this paper we proposed a deep reinforcement learning framework D²RLIR to model an interactive recommendation system, which generates the recommendation list concerning both relevancy and diversity. Unlike previous works, which had redundant computation in generating the recommendations, we propose to use an approximate nearest neighbor search (ANNS) method to find the closest items to the proto-action and then generate a diverse list using the TDE algorithm. The results of experiments show that it is not necessary to use complicated structures such as RNNs and CNNs to construct the user state, and the user state can be constructed using positional encoding that has a low computational complexity. The use of positional encoding does not only reduce the precision and diversity of the recommendations list, but it also gives better performance than other methods.

As future work, we intend to examine the effect of adding self-attention mechanism and combining it with positional encoding on the precision and diversity of the generated recommendations list. Additionally, we are interested in using probability distributions instead of offline datasets such as MovieLens to train algorithm.